\documentstyle[emulateapj,psfig]{article}  
\slugcomment{accepted for publication in ApJ Letter}  
\lefthead{ZHANG \& HARDING}  
\righthead{HIGH B PULSARS AND MAGNETARS: A UNIFIED PICTURE}  
   
\begin{document}   
\title{High magnetic field pulsars and magnetars: a unified picture}
\author{Bing Zhang\altaffilmark{1}
 and Alice K. Harding}   
\affil {Laboratory of High Energy Astrophysics, NASA Goddard Space  
 Flight Center, Greenbelt, MD 20771\\ 
 bzhang@twinkie.gsfc.nasa.gov, harding@twinkie.gsfc.nasa.gov }
\altaffiltext{1}{National Research Council Research Associate Fellow.}  
  
\begin{abstract}
We propose a unified picture of high magnetic field radio pulsars and 
magnetars by arguing that they are all rotating high-field neutron 
stars, but have different orientations of their magnetic axes with 
respective to their rotation axes. In strong magnetic fields where 
photon splitting suppresses pair creation near the surface, the 
high-field pulsars can have active inner accelerators while the 
anomalous X-ray pulsars cannot. This can account for the very different 
observed emission characteristics of the anomalous X-ray pulsar 1E 
2259+586 and the high field radio pulsar PSR J1814-1744. A predicted 
consequence of this picture is that radio pulsars having surface 
magnetic field greater than about $2\times 10^{14}$ G should not exist. 
\end{abstract}   
   
\keywords{stars: magnetic fields - stars: neutron - pulsars: general - 
 pulsars: individual: PSR J1814-1744, 1E 2259+586}
  
\section{Introduction}
There is growing evidence that two sub-groups of objects, namely
soft $\gamma$-ray repeaters (hereafter SGRs) and anomalous X-ray 
pulsars (hereafter AXPs) are magnetars (e.g. Kouveliotou et al. 1998, 
1999; Hurley et al. 1999; Mereghetti \& Stella 1995; Wilson et al. 1999; 
Kaspi, Chakrabarty \& Steinberger 1999), a type of objects with dipolar 
magnetic fields much stronger than the critical magnetic field (Duncan 
\& Thompson 1992; Paczynski 1992; Usov 1992; Thompson \& Duncan 1995, 
1996). These objects occupy a unique phase space in their combination 
of long, monotonically increasing periods and high period derivatives, 
and are believed to be a distinct species from the normal radio pulsars 
in that most of them are radio quiet, except for a possible detection 
of radio emission from SGR 1900+14 (Shitov 1999; Shitov, Pugachev \& 
Kutuzov 2000). However, the recent Parkes multi-beam radio pulsar survey 
(e.g. Manchester et al. 2000; Camilo et al. 2000b) discovered three 
pulsars with dipolar field strength higher than the critical value (i.e. 
high magnetic field pulsars, hereafter HBPs); and one of them, PSR 
J1814-1744, has spin parameters quite similar to the AXP 1E 2259+586 
(Camilo et al. 2000; Kaspi et al. 2000). Furthermore, a search of the 
archival X-ray data from the HBP PSR J1814-1744 indicates that the upper 
limit of the X-ray luminosity of this pulsar is approximately 1/10 that 
of 1E 2259+586; this led to the suggestion that HBPs and AXPs may have 
distinct evolutionary paths, despite their proximity in period-period 
derivative phase space (Pivoraroff, Kaspi \& Camilo 2000). 
Here we propose a possible interpretation of the distinct emission
properties of the HBPs and AXPs (especially PSR J1814-1744 and 1E 
2259+586) using a simple geometric effect. 

\section{The interpretation}

Pulsar radio emission is believed to be due to some kind of coherent 
emission processes in an electron-positron pair plasma. Therefore pair 
production from a pulsar inner magnetosphere is the essential condition 
for pulsar radio emission. The apparent lack of pulsed radio emission from
the known SGRs and AXPs has been attributed to the possible pair production
suppression by photon splitting, a third-order QED process that may become 
important in strong magnetic fields above the critical field 
(Baring \& Harding 1998, 2000). However, the discovery of 
PSR J1119-6127, PSR J1726-3530, and PSR J1814-1744 (Camilo et al. 2000a;
Kaspi et al. 2000, see also, http://www.atnf.csiro.au/$\sim$pulsar
/psr/pmsurv/pmwww/pmpsrs.db) above the photon splitting ``death line''
(Baring \& Harding 1998) raises questions concerning how photon 
splitting suppresses pair production. The close clustering of PSR 
J1814-1744 and the AXP 1E 2259+586 in the $\dot P-P$ phase space makes
the problem more severe.

Such a behavior could be understood in a simple, unified picture
when the properties of pulsar inner accelerators are taken into account,
assuming that photon splitting can completely suppress pair creation in
magnetic fields exceeding $\sim 10^{14}$ G. Baring \& Harding (2000) 
found that pair creation is completely suppressed only if all three 
modes permitted by QED operate, i.e. photons polarized both parallel and
perpendicular to the field can split, which we will assume throughout 
this letter. Rotating magnetized neutron stars are unipolar inductors 
that generate huge potential drops
\begin{equation}
\Phi \sim {B_p R^3 \Omega^2 \over 2c^2} =(1.0\times 10^{13} {\rm V})
\left({B_p \over 10^{14}{\rm G}}\right)\left({P\over 8{\rm s}}\right)
^{-2}R_6^3
\end{equation}
across the open field line region, where $B_p$ is the dipolar magnetic
field strength at the pole, $R$ is the star radius, and $\Omega$, $P$ 
are the rotation velocity and the period of the pulsar, respectively. 
Under certain conditions a part of or even the total amount of this 
potential will drop across a charge-depleted region (or a gap) formed 
in the polar cap area of the pulsar. Depending on the boundary condition 
at the surface, there are two kinds of inner accelerators, i.e.,
the space-charge-limited flow (hereafter SCLF) type accelerator, which 
is formed by a self-regulated flow due to free extraction of the charged 
particles from the neutron star surface (Arons \& Scharlemann 1979; 
Harding \& Muslimov 1998), and the vacuum (hereafter V) type gap, which 
is formed due to strong binding of the charged particles within the 
surface (Ruderman \& Sutherland 1975).
There are two important differences between these two types of 
accelerators (see Zhang, Harding \& Muslimov 2000, for a comparison 
between the two models). First, SCLF gaps could be extremely long and 
narrow because the potential drop only increases mildly with the gap
height so that pair formation fronts (PFFs) could be formed at much 
higher altitudes, especially
when the parallel electric field within the gap is saturated 
(Harding \& Muslimov 1998; Zhang \& Harding 2000). 
A V gap, on the other hand, usually has a height less than the 
polar cap radius mainly due to the quadratic form of the gap 
potential-height relation, and the maximum potential is achieved when 
the gap height $h\sim r_{\rm pc}/\sqrt{2}$, where $r_{\rm pc}$ is the 
polar cap radius (Ruderman \& Sutherland 1975; Zhang et al. 2000). 
Second, the cessation of a V gap requires a pair formation avalanche 
within the gap, which requires pair production at both the top and the 
bottom of the gap (Ruderman \& Sutherland 1975). A SCLF gap, however, 
only requires a steady PFF at the top of the gap\footnote{Although a 
lower PFF at the gap bottom can also be formed sometimes in the SCLF 
model (e.g. Harding \& Muslimov 1998), this is never obligatory.}. 

These features have important implications for whether photon splitting 
can completely suppress pair production. Since photon splitting is only 
important in strong fields, it does not suppress pair production at higher 
altitudes where the local field strength is considerably degraded ($B(r)
\propto r^{-3}$, where $r$ is the radius in spherical coordinates), even 
if the near surface field is super-strong. Since a SCLF accelerator
could be long and narrow, particles within the gap can keep 
accelerating. Though the $\gamma$-rays produced near the surface by these 
primary particles will split to lower frequency photons, 
those produced at higher altitudes will eventually undergo pair production 
in the less intense fields. For a V gap, however, the gap is usually 
pancake-shaped, and there is no significant degradation of the field at 
the gap top with respect to the gap bottom. Even if in some cases (usually 
near the death line, Zhang et al. 2000) the gap could be long so that the 
field at the top is less intense, the gap still can not breakdown because 
pairs can not be formed at the bottom. Thus if the surface field strength 
is super strong, very likely, such a V gap simply does not form, since a 
gap solution with boundary condition [$E_\parallel(z=0)\neq 0$, $E_\parallel
(z=h)=0$] can not be realized. We therefore conclude that {\em in magnetar 
environments where photon splitting effectively suppresses pair production 
at the surface, only SCLF-type accelerators could be formed; V-type 
accelerators can not develop.} 
Numerical simulations (Baring \& Harding 2000) show that the pair yields 
drop steeply with increasing magnetic field when photon splitting starts 
to suppress pair production, which means that at the top of the SCLF gap 
when pair production starts to overcome photon splitting the pair 
production rate also rises steeply to provide copious pairs for radio 
emission. Here we assume that {\em in magnetar environments, radio emission 
is possible if a SCLF accelerator is formed.}

We now discuss the condition for a SCLF gap to develop. The essential
point is to investigate whether free ejection of charged particles 
from the surface is possible, and this depends on (1) the binding 
energy of the charged particles, and (2) the surface temperature. Let 
us first discuss a type of neutron star in which the angle between 
the rotation axis and the magnetic axis is larger than $90^{\rm o}$, 
or ${\bf \Omega \cdot B}_p <0$. We call such rotators ``anti-parallel 
rotators'' (hereafter APRs) although the axes are not strictly 
anti-parallel. In such rotators, a force-free magnetosphere requires 
that the polar cap region is filled with positive charges, and the 
positive ions from the surface are expected to flow out. The 
composition of the neutron star surface is uncertain, but it is 
probably composed of $^{56}$Fe. Although there is a large uncertainty 
in calculating the binding energy of Fe ions, in magnetar environment, 
it is likely that a magnetic metal could be formed, and the cohesive 
energy could be approximated $\Delta \epsilon \simeq (26.0{\rm keV})
(B_p /10^{14}{\rm G})^{0.73}$ (Abrahams \& Shapiro 1991; Usov \&
Melrose 1995, 1996), and the critical temperature for thermionic 
emission of the ions is (Usov \& Melrose 1995)
\begin{equation}
T_i\simeq (1.0\times 10^7{\rm K})\left({B_p \over 10^{14}{\rm G}}
\right)^{0.73}.
\end{equation}
The surface temperature of a magnetar is connected to the core 
temperature by, e.g., $T_s \simeq (0.87\times 10^6 {\rm K})(T_c/10^8
{\rm K})^{0.55}g_{14}^{0.25}$ (Gudmundsson, Pethick \& Epstein 1983,
where $g_{14}$ is the surface gravity of the neutron star in units 
of $10^{14}$ cm s$^{-2}$). The balance of neutrino cooling 
and various magnetic heating yields the core temperature to be $T_c 
\sim 6\times 10^8$ K which insensitively depends on the age of the 
magnetar (Thompson \& Duncan 1996). This approach gives a rough estimate 
of the surface temperature of $T_s \sim 2.3 \times 10^6$ K. From the 
observational approach, the surface temperature could be estimated from 
the quiescent X-ray luminosity of magnetars, $L_x$, by 
\begin{equation}
T_s\simeq (4.8\times 10^6 {\rm K}) \left({L_x \over 10^{35}{\rm 
erg~s^{-1}}}\right)^{1/4} \left({R\over 10{\rm km}}\right)^{-1/2}.
\end{equation}
Both estimations show that the surface temperature is not hot enough 
for thermionic emission of the ions, and that a SCLF accelerator can 
not develop. With $T_s\sim 3\times 10^6$ K, the condition for a SCLF
gap, i.e., $T_s\geq T_i$, is $B_p^i \leq 1.9\times 10^{13}$ G, which is 
well below the photon splitting death line for surface emission (Baring 
\& Harding 1998)
\begin{equation}
B_p \simeq (5.7\times 10^{13} {\rm G}) P^{2/15}.
\label{photonsplitting}
\end{equation}
Field emission of ions, which is possibly important when thermionic 
emission is unimportant, can also be neglected, since the maximum 
parallel electric field at the surface $E_\parallel({\rm max})
=(2\Omega B_p/c)(r_{\rm pc}/\sqrt{2})\sim 5.7\times 10^9 
({\rm V~cm^{-1}})B_{p,14}(P/8{\rm s})^{-3/2}(R/10 {\rm km})
^{3/2}$ is much smaller than the critical parallel field required to 
pull out the ions, $E_\parallel({\rm cri})\simeq (8\times 10^{12}
{\rm V~cm^{-1}})(\Delta\epsilon / 26{\rm keV})^{3/2}$ (Usov \& Melrose
1995). Here $r_{\rm pc}\sim R(\Omega R/c)^{1/2}$ is the radius of the 
polar cap, and $B_{p,14}=B_p/(10^{14}{\rm G})$. 
Therefore, for APRs above the photon splitting death line 
(\ref{photonsplitting}), the magnetospheres of magnetars are very likely 
to be dead, with no active inner accelerators.

\centerline{}
\centerline{\psfig{file=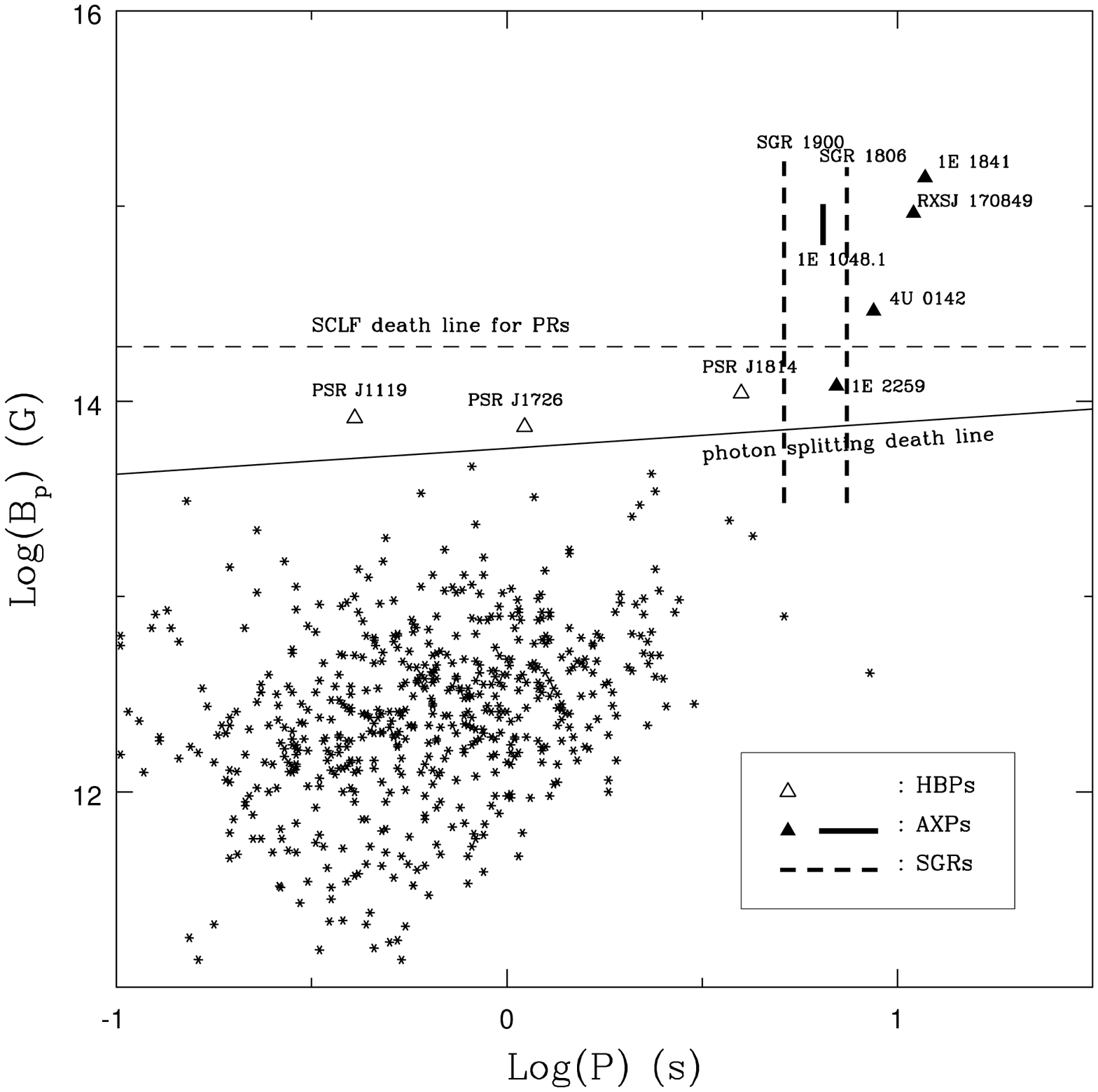,width=10.0cm}}
\figcaption{A $B_p-P$ diagram of some radio pulsars and known magnetars. 
The spin parameters of the HBPs are from Kaspi et al. (2000) and
http://www.atnf.csiro.au/ $\sim$pulsar/ psr/ pmsurv/ pmwww/ pmpsrs.db. The
spin parameters of the AXPs and SGBs are from Mereghetti \& Stella
(1995), Wilson et al. (1999), Mereghetti (1999), and Kouveliotou et
al. (1998). Triangles mean precise measurements of unvarying $\dot
P$. Variations of $\dot P$ data for AXP 1E 1048.1-5937 follows
Mereghetti (1999) and Paul et al. (2000). The upper ends of the two
bars for the two SGRs are estimated by assuming pure dipolar
spin-down, while the lower ends are estimated by assuming that
spin-down is dominated by a continuous wind of luminosity 
$10^{37} {\rm erg~s^{-1}}$ following Harding et al. (1999). 
The solid line is the photon splitting death line for surface emission 
(Baring \& Harding 1998), which acts as the radio emission death line of 
the APRs. The dashed line (SCLF death line for PRs) acts as the radio 
emission death line for the PRs.}
\centerline{}

We now investigate the opposite case that the neutron stars have
inclinations less than $90^{\rm o}$, i.e., ${\bf \Omega \cdot B}_p 
>0$. With the same convention, we call such kind of objects parallel 
rotators (hereafter PRs). In such neutron stars, negative charges are 
expected to fill the polar cap region, and electrons from the surface 
are expected to be pulled out. The binding energy of electrons is 
roughly the Fermi energy of the electrons, which reads 
$\epsilon_{\rm F}\simeq (5.0 {\rm keV})(Z/26)^{0.8}B_{p,14}^{0.4}$
(Usov \& Melrose 1995). And the critical temperature for thermionic 
emission of the electrons is
\begin{equation}
T_e \simeq (2.3\times 10^6 {\rm K})\left({Z\over 26}\right)^{0.8} 
\left({B_p\over 10^{14}{\rm G}}\right)^{0.4},
\end{equation}
where $Z$ is the atomic number ($Z=26$ for Fe) (Usov \& Melrose 1995). 
Assuming that the fields of the HBPs decay in a similar way to that of 
the AXPs, they should also have typical surface temperatures of $T_s 
\sim 3\times 10^6$ K. By comparing the temperature $T_e$ to $T_s$ in 
magnetar environments, the condition for thermionic emission of electrons 
(and hence for a SCLF accelerator to form for a PR), i.e., $T_s \geq T_e$, 
is thus
\begin{equation}
B_p^e \lesssim 1.9 \times 10^{14} {\rm G}.
\label{sclf}
\end{equation}
It is notable that thermionic emission of the electrons is possible at 
a much higher field strength than the field strength $B_p^i$ 
for the thermionic emission of ions.  There is a large 
phase space in the high $B$ regime where pair production is still 
allowed at higher altitudes when a SCLF accelerator is formed, even 
though it is above the photon splitting death line for surface emission. 
We emphasize that such a conclusion only holds for PRs, and it is 
interesting to note that such rotators are originally proposed as 
``anti-pulsars'' in the Ruderman \& Sutherland (1975) vacuum gap model. 
 
A $B_p-P$ diagram of the HBPs, AXPs, SGRs, as well as some of the
other known radio pulsars is shown in Fig.1, in which the photon
splitting death line and the SCLF accelerator ``death line'' for PRs
are also plotted. Polar surface magnetic field is calculated by 
$B_p =6.4\times 10^{19}\sqrt{P\dot P}$ G (Shapiro \& Teukolsky 1983;
Usov \& Melrose 1995). Note that besides  
the observed variation of $\dot P$ for SGR 1900+14 and 1E 1048.1-5937, 
there are even more uncertainties in inferring $B_p$ from the spin-down 
parameters of SGRs, since a general spin-down formula also includes the 
contribution from the winds (Harding, Contopoulos, \& Kazanas 1999). 
Therefore, in some cases we adopt a bar instead of a point to denote 
an object in Fig.1. An important fact is that the three newly-discovered 
HBPs are located in the phase space defined by the photon splitting and
PR SCLF death lines, which means that they could be theoretically radio 
loud if they are PRs. The AXP 1E 2259+586 also lies in 
this regime but is clearly radio quiet, and we argue that it is an APR. 
Thus the discrepancy between AXPs and HBPs in this picture is simply due 
to a geometric effect.

Another question is why PSR J1814-1744 is quiet in X-rays, in contrast
to 1E 2259+586. Quiescent emission from magnetars is interpreted as due
to magnetic field decay (Goldreich \& Reisenegger \& 1992; Thompson \& 
Duncan 1996; Heyl \& Kulkarni 1998). The quiescent X-ray luminosity
should satisfy 
\begin{equation}
L_x\leq \dot E_B \simeq (1/6)B_p^2 R^3/\tau_d,
\label{Lx}
\end{equation}
where $\dot E_B$ is the magnetic energy decay rate, and $\tau_d$ is the 
decay time scale, which is model dependent (Goldreich \& Reisenegger
1992; Heyl \& Kulkarni 1998) and $B$ dependent. Note that $E_B=(1/12)
B_p^2 R^3$, and $\dot B\sim B/\tau_d$ have been adopted. In the general 
case of a field permeating the core, ambipolar diffusion is the dominant 
decay mechanism when the field strength is in the magnetar regime. For 
the solenoidal ambipolar decay mode, the decay timescale is $\tau_d({\rm 
ambip,s})\sim (3\times 10^5 {\rm yr})L_5^2 T_{c,8}^2 B_{p,14}^2$, 
thus the quiescent X-ray luminosity is limited to
\begin{equation}
L_x({\rm ambip,s}) \lesssim 1.8\times 10^{32} {\rm ergs~s^{-1}} 
B_{p,14}^4 R_6^3 L_5^{-2} T_{c,8}^{-2}.
\label{decay}
\end{equation}
Here $T_{c,8}=T_c/10^8$ K, $R_6=R/10^6$ cm, and $L_5$ is a characteristic 
length scale of the flux loops through the outer core in units of $10^5$ 
cm (Goldreich \& Reisenegger 1992). Note that the right side is 
proportional to $B_p^4$. Observed quiescent X-ray luminosities of all 
other AXPs do not contradict (\ref{decay}) except for 1E 2259+586, 
which gives an upper limit on the X-ray luminosity of $L_x({\rm 1E2259}) 
\lesssim 3.7\times 10^{32} {\rm ergs~s^{-1}}$, while observations show that 
$L_x({\rm 1E2259,obs}) \sim 5\times 10^{34}{\rm ergs~s^{-1}}$. For PSR 
J1814-1744, on the other hand, (\ref{decay}) gives $L_x ({\rm J1814})
\lesssim 2.6\times 10^{32} {\rm ergs~s^{-1}}$, which is consistent with the 
upper limit from the archival observations, $L_x ({\rm J1814,obs}) < 3.8
\times10^{33} {\rm ergs~s^{-1}}$ [$\sim (1/13)L_x ({\rm 1E2259,obs})$, 
Pivovaroff et al. 2000]\footnote{Note that this luminosity upper limit 
implies a surface 
temperature above what we have assumed to derive the SCLF death line for 
PRs (eq.[\ref{sclf}]).}. Thus the non-detection of bright X-rays from PSR 
J1814-1744 is not a surprise if neutron star magnetic fields are of 
core-origin. The discrepency between PSR J1814-1744 and 1E 2259+586 must 
then be attributed to the peculiarity of 1E 2259+586. To interpret the 
quiescent emission of this AXP, one needs to assume either a much faster 
field decay mechanism (e.g. magnetic fields are of crustal origin and the 
Hall cascade effect dominates the field decay, Colpi, Geppert \& Page
2000) or much stronger multipole fields near the magnetic pole (Baring
\& Harding 2000). In fact, the age of the associated supernova remnant 
(SNR) CTB 109 ($t_{\rm SNR}\sim 3-20$ kyr) (Parmar et al. 1998) is much 
younger than the characteristic age of the AXP 1E 2259+586 ($t_c \sim 
2.3\times 10^5$ yr) and $\tau_d({\rm ambip,s})$. If one assumes $\tau_d 
\sim t_{\rm SNR}$ for this pulsar, the high $L_x$ is then consistent 
with eq.(\ref{Lx}).

\section{Discussions}

In this letter
we discuss possible formation of the inner accelerators in a magnetar
environment for the first time and come to a unified picture for AXPs 
and HBPs by arguing that they are all rotating high-field 
neutron stars, but have different orientations of the magnetic axes 
with respective to the rotation axes. If photon splitting suppresses
pair creation near the surface, the HBPs can have active inner 
accelerators while the AXPs can not. 

This suggestion may also have implications for
another type of magnetar, i.e., the SGRs. These objects react much
differently from the AXPs by exhibiting irregular short bursts and 
occasional giant flares, which are interpreted as crust cracking and
large-scale magnetic field reconnection, respectively, within the 
framework of the magnetar model (Thompson \& Duncan 1995). It remains 
unclear whether they are experiencing a different evolutionary stage 
than that of AXPs or whether they are intrinsically different objects. 
Only two of them (SGR 1806-20 and SGR 1900+14) have $\dot P$ 
measurements, but determination of their dipolar magnetic fields is
complicated by the contribution of the relativistic winds to the
spin-down (Harding et al. 1999). It is notable that the constraints of 
both the SNR age and the magnetar energy requirements lead to a polar 
field $B_p({\rm SGR})\sim 10^{14}$ G (Harding et al. 1999), which may lie 
below the SCLF death line for PRs. Thus they may also have active inner 
accelerators if they are actually PRs. Here we suggest a possibility that 
SGRs might also have active accelerators while AXPs do not, and the active 
behaviors of SGRs may have some connections with their inner accelerators. 
For example, the constant extraction of electrons from the pole may 
somehow more frequently trigger instability within the crust. 
One expectation of this scenario is pulsed radio emission from the SGRs, 
which may account for the pulsed radio emission from SGR 1900+14 (Shitov 
1999; Shitov et al. 2000).

Theoretically, the question of whether photon splitting occurs in all 
three modes permitted by QED or only in one mode in superstrong magnetic 
fields is difficult to tackle. If radio pulsar surveys discover any 
pulsar above the SCLF death line for PRs (the dashed line in Fig.1), 
these pulsars must have active V gaps with pair breakdown near the
surface. This would strongly imply that only one mode of photon 
splitting occurs in fields above a few times $10^{14}$ G, and
thus sheds important light on a fundamental physics process. It is 
worth noting that the location of the SCLF death line for PRs depends 
on the surface temperature of the neutron star, so that the location 
of the dashed line in Fig.1 may rise or drop. Thus detections in both 
radio and X-ray bands are desirable.

We thank Matthew Baring, Alex Muslimov, and Zaven Arzoumanian 
for interesting discussions and helpful comments.




\end{document}